\documentclass[12pt,english,a4paper,dvips]{article}
\usepackage[T1]{fontenc}
\usepackage[utf8]{inputenc}
\usepackage{amsmath}

\makeatletter

\usepackage{mathptmx}

\usepackage[scaled]{helvet}

\usepackage{courier}

\usepackage{ucs}

\usepackage{babel}
\makeatother
\begin{document}

\title{On the origin dependence of multipole moments in electromagnetism.}

\author{Patrick De Visschere\\
\\
Ghent University, Dept. ELIS\\
Sint-Pietersnieuwstraat 41, B-9000 Gent, Belgium\\
tel. +32-9-264 8950; fax. +32-9-264 35 94; e-mail: pdv@elis.UGent.be}

\maketitle
\begin{abstract}
The standard description of material media in electromagnetism is
based on multipoles. It is well known that these moments depend on
the point of reference chosen, except for the lowest order. It is
shown that this “origin dependence” is not unphysical as has been
claimed in the literature but forms only part of the effect of moving
the point of reference. When also the complementary part is taken
into account then different points of reference lead to different
but equivalent descriptions of the same physical reality. This is
shown at the microscopic as well as at the macroscopic level. A similar
interpretation is valid regarding the “origin dependence” of the
reflection coefficients for reflection on a semi infinite medium.
We show that the “transformation theory” which has been proposed
to remedy this situation (and which is thus not needed) is unphysical
since the transformation considered does not leave the boundary conditions
invariant. 
\end{abstract}

\section{Introduction}

In classical electrodynamics\cite{CHDFAAEE} material media are modelled
at the microscopic level as an ensemble of stable building blocks
(atoms, ions, molecules, …) with certain microscopic charge ($\eta$)
and current ($\overline{j}$) densities. The microscopic fields ($\overline{e},\overline{b}$)
then obey the \emph{microscopic} Maxwell equations: \begin{equation}
\begin{split}\nabla\times\overline{e}=-\frac{\partial\overline{b}}{\partial t}\phantom{\rule{10mm}{0ex}}\nabla\cdot\overline{b}=0\\
\nabla\times\frac{\overline{b}}{{\mu}_{0}}=\frac{\partial{\epsilon}_{0}\overline{e}}{\partial t}+\overline{j}\phantom{\rule{10mm}{0ex}}\nabla\cdot{\epsilon}_{0}\overline{e}=\eta\end{split}
\label{eq:CHDHCGJA}\end{equation}

With a purely classical model the microscopic sources may be written
as: \begin{equation}
\begin{split}\eta=\sum_{k,l}{q}_{kl}\delta(\overline{r}-{\overline{R}}_{kl})\phantom{\rule{10mm}{0ex}}\overline{j}=\sum_{k,l}{q}_{kl}{\dot{\overline{R}}}_{kl}\delta(\overline{r}-{\overline{R}}_{kl})\end{split}
\label{eq:CHDFDBBB}\end{equation}

where the index \emph{k} labels the building blocks and \emph{l} the
point charges (with position ${\overline{R}}_{kl}$ and velocity ${\dot{\overline{R}}}_{kl}$)
within a building block. The \emph{macroscopic} Maxwell equations
then follow by a suitable averaging procedure. Due to the linearity
these equations are similar to the ones in (\ref{eq:CHDHCGJA}) with
($\overline{e},\overline{b}$) replaced by the macroscopic fields
($\overline{E},\overline{B}$) and with the microscopic sources replaced
by their averages ($\langle\eta\rangle,\langle\overline{j}\rangle$).
The microscopic source densities (\ref{eq:CHDFDBBB}) can be rearranged
and expressed in terms of the (microscopic) electric and magnetic
multipole moment densities. Usually only the 2 lowest order multipole
moments are taking into account. After averaging, these lead to the
usual macroscopic source densities: \begin{equation}
\begin{split}\langle\eta\rangle=\rho-\nabla\cdot\overline{P}\\
\langle\overline{j}\rangle=\overline{J}+\frac{\partial\overline{P}}{\partial t}+\nabla\times\overline{M}\end{split}
\label{eq:CHDGCJDC}\end{equation}

where ($\rho,\overline{J}$) are the unbound charge and current densities
and $\overline{P}$ and $\overline{M}$ the polarization and magnetization
densities. In order to avoid the complexities associated with moving
media, in what follows we will consider only non-moving media. Substituting
(\ref{eq:CHDGCJDC}) in the averaged (\ref{eq:CHDHCGJA}) and defining
$\overline{D}={\epsilon}_{0}\overline{E}+\overline{P}$ and $\overline{H}=\frac{\overline{B}}{{\mu}_{0}}-\overline{M}$
one finds the standard macroscopic Maxwell equations: \begin{equation}
\begin{split}\nabla\times\overline{E}=-\frac{\partial\overline{B}}{\partial t}\phantom{\rule{10mm}{0ex}}\nabla\cdot\overline{B}=0\\
\nabla\times\overline{H}=\frac{\partial\overline{D}}{\partial t}+\overline{J}\phantom{\rule{10mm}{0ex}}\nabla\cdot\overline{D}=\rho\end{split}
\label{eq:CHDBAFJB}\end{equation}

If higher order multipole contributions are taken into account then
additional terms show up in (\ref{eq:CHDGCJDC}). By adapting the
definition of ($\overline{D},\overline{H}$) accordingly equations
(\ref{eq:CHDBAFJB}) remain valid. However it should be noted that
at a boundary between 2 media the jump conditions which follow from
(\ref{eq:CHDBAFJB}) have to be adapted in order to take into account
these additional multipole contributions.

It is well known that the multipole moments are not unique and more
precisely they depend on the choice of the reference point (e.g. the
centre of mass) used for calculating the moments, except for the lowest
order non vanishing moment. E.g. the total charge of an ion ${q}_{k}=\sum_{l}{q}_{kl}$
is independent of the reference point but the electric dipole moment
is not, except if ${q}_{k}=0$. For insulating materials the dipole
terms are unique and since higher moments are usually not considered
the non-uniqueness of the higher moments is often of no consequence.

In \cite{CHDFBJJA} \cite{CHDBJBCB} material media are treated up
to the electric octopole and the magnetic quadrupole terms. The emphasis
is on optical phenomena (in insulating media) and in particular Fresnel
reflection coefficients are calculated for a semi-infinite medium
taking into account these quadrupole and octopole effects. Therefore
the authors are faced with the problem of the non-uniqueness of these
higher order moments. As a result of lengthy calculations they find
that in general those reflection coefficients also depend on the chosen
reference point (it is called “origin dependent”, but this should
not be mistaken for the origin of the laboratory frame, but as a reference
point within an atom.) These results are then discarded as being “unphysical”
and using a “transformation theory” they adapt the “standard
theory” so that “origin-independent” reflection coefficients
are obtained. Although at first sight the reasoning may seem tempting
no indication is given at what point exactly the “standard theory”
became “unphysical”. After some consideration we became convinced
that there is actually no problem with the “standard theory” and
that on the contrary the “origin dependence” as calculated in
\cite{CHDFBJJA} has no physical meaning since only part of the effect
of changing the position of the “origin” has been taken into account.
In fact when choosing another reference point not only the higher
order multipole moments do change but so does the position of these
elementary sources. If both effects are taken into account then the
microscopic charge density remains unchanged as expected. This is
explained in more detail in section 2. In sections 3 and 4 the same
idea is worked out in the macroscopic domain and in section 5 the
effect on reflection coefficients is considered. It will become clear
that there is no need for a “transformation theory”. On the contrary,
since the “standard theory” and the “transformation theory”
give different results, at least one of them must be unphysical. In
the last section 6 it will be shown that the “transformation theory”
is unphysical since it does not leave the boundary conditions invariant.

\section{Microscopic charge and current densities}

In order to simplify the notation we consider only a single building
block. Then up to quadrupole order the microscopic charge density
is written as: \begin{equation}
\begin{split}\eta=q\delta(\overline{r}-\overline{R})-{p}_{i}{\partial}_{i}\delta(\overline{r}-\overline{R})+{q}_{ij}{\partial}_{ij}\delta(\overline{r}-\overline{R})+\cdots\end{split}
\label{eq:CHDCICJA}\end{equation}

where the electric multipole moments are defined as\cite{CHDEIBID}:
\begin{equation}
\begin{split}\mu^{\left(n\right)}=\frac{1}{n!}\sum_{l}{q}_{l}{\overline{r}}_{l}^{n}\end{split}
\label{eq:}\end{equation}

In (\ref{eq:CHDCICJA}) we use $\left(q,{p}_{i},{q}_{ij}\right)$
as a shorthand notation for the 3 lowest order multipole moments,
${\partial}_{i}$ for the derivative with respect to ${x}_{i}$, the
cartesian components of $\overline{r}$, $\overline{R}$ for the position
vector of the reference point and ${\overline{r}}_{l}$ for the position
vector of charge \emph{l} with respect to this reference point (thus
${\overline{R}}_{l}=\overline{R}+{\overline{r}}_{l}$, where compared
with (\ref{eq:CHDFDBBB}) the index \emph{k} has been omitted). If
we choose another reference point within the building block via $\overline{R}={\overline{R}}^{\prime}+\overline{d}$
and ${\overline{r}}_{l}={{\overline{r}}^{\prime}}_{l}-\overline{d}$
then in general different multipole moments are obtained: \begin{equation}
\begin{split}{p}_{i}=\sum_{l}{q}_{l}{x}_{il}=\sum_{l}{q}_{l}({{x}^{\prime}}_{il}-{d}_{i})={{p}^{\prime}}_{i}-q{d}_{i}\end{split}
\label{eq:CHDDAFGG}\end{equation}

and similarly for the quadrupole moment: \begin{equation}
\begin{split}{q}_{ij}={{q}^{\prime}}_{ij}-\frac{1}{2}({d}_{i}{{p}^{\prime}}_{j}+{{p}^{\prime}}_{i}{d}_{j})+\frac{1}{2}q{d}_{i}{d}_{j}\end{split}
\label{eq:CHDHJECI}\end{equation}

On the other hand these modified moments must now be placed in the
new reference point. With a Taylor series we write: \begin{equation}
\begin{split}\delta(\overline{r}-\overline{R})=\delta(\overline{r}-{\overline{R}}^{\prime}-\overline{d})\\
\phantom{x}=\delta(\overline{r}-{\overline{R}}^{\prime})-{d}_{i}{\partial}_{i}\delta(\overline{r}-{\overline{R}}^{\prime})+\frac{1}{2}{d}_{i}{d}_{j}{\partial}_{ij}\delta(\overline{r}-{\overline{R}}^{\prime})+\cdots\end{split}
\label{eq:CHDFCDEB}\end{equation}

Inserting (\ref{eq:CHDDAFGG}) (\ref{eq:CHDHJECI}) and (\ref{eq:CHDFCDEB})
into (\ref{eq:CHDCICJA}) and retaining only terms up to quadrupole
order one finds a similar expression but referred to the new reference
point: \begin{equation}
\begin{split}\eta=q\delta(\overline{r}-{\overline{R}}^{\prime})-{{p}^{\prime}}_{i}{\partial}_{i}\delta(\overline{r}-{\overline{R}}^{\prime})+{{q}^{\prime}}_{ij}{\partial}_{ij}\delta(\overline{r}-{\overline{R}}^{\prime})+\cdots\end{split}
\label{eq:}\end{equation}

Therefore the microscopic charge density ($\eta$) is as expected
independent of the choice of reference point within the building block,
although the multipole moments themselves do change according to (\ref{eq:CHDDAFGG})
and (\ref{eq:CHDHJECI}). The interpretation is straightforward: suppose
$q\neq0$ then moving the reference point over $-\overline{d}$ changes
the dipole moment with $q\overline{d}$ but the monopole \emph{q}
must be moved with the reference point and this constitutes an additional
dipole moment exactly cancelling the first change.

A similar conclusion can be drawn with respect to the microscopic
current density. We define the magnetic multipole moments by: \begin{equation}
\begin{split}\nu^{\left(n\right)}=\frac{n}{(n+1)!}\sum_{l}{q}_{l}{\overline{r}}_{l}^{n-1}({\overline{r}}_{l}\times{\dot{\overline{r}}}_{l})\end{split}
\label{eq:}\end{equation}

with in particular for the magnetic dipole moment ($n=1$): \begin{equation}
\begin{split}{m}_{i}=\frac{1}{2}\sum_{l}{q}_{l}{e}_{ijk}{x}_{jl}{\dot{x}}_{kl}\end{split}
\label{eq:}\end{equation}

where ${e}_{ijk}$ is the Levi-Civita tensor. This dipole moment transforms
according to: \begin{equation}
\begin{split}{m}_{i}={{m}^{\prime}}_{i}-\frac{1}{2}{e}_{ijk}{d}_{j}{{\dot{p}}^{\prime}}_{k}\end{split}
\label{eq:CHDBAIJC}\end{equation}

Up to electric quadrupole and magnetic dipole order, the microscopic
current density is then given by: \begin{equation}
\begin{split}{j}_{i}={\dot{p}}_{i}\delta(\overline{r}-\overline{R})-{\dot{q}}_{ji}{\partial}_{j}\delta(\overline{r}-\overline{R})+{e}_{ijk}{m}_{k}{\partial}_{j}\delta(\overline{r}-\overline{R})\end{split}
\label{eq:CHDHHJBI}\end{equation}

Substituting (\ref{eq:CHDDAFGG}) (\ref{eq:CHDHJECI}) (\ref{eq:CHDBAIJC})
and (\ref{eq:CHDFCDEB}) into (\ref{eq:CHDHHJBI}) again results in
a similar expression for the microscopic current density with respect
to the new reference point: \begin{equation}
\begin{split}{j}_{i}={{\dot{p}}^{\prime}}_{i}\delta(\overline{r}-{\overline{R}}^{\prime})-{{\dot{q}}^{\prime}}_{ji}{\partial}_{j}\delta(\overline{r}-{\overline{R}}^{\prime})+{e}_{ijk}{{m}^{\prime}}_{k}{\partial}_{j}\delta(\overline{r}-{\overline{R}}^{\prime})\end{split}
\label{eq:}\end{equation}

We consider now as an example the reflection of a plane wave impinging
on a semi infinite medium as was treated in \cite{CHDFBJJA}. In principle
the problem could be solved at the microscopic level and this would
result in definite values for the charge and current densities in
each atom of the medium and thus for the global $\eta$ and $\overline{j}$.
We could then describe these resulting charge and current densities
using multipole moments up to some chosen order and as shown this
representation is independent of the choice of reference point, although
the moments themselves are “origin dependent”. This shows that
already at the microscopic level the “origin dependence” of the
multipole moments has no physical consequence supposed they are treated
in a consequent way, meaning that the multipoles should be placed
at their particular reference point.

\section{Macroscopic charge and current densities}

Since the macroscopic charge and current densities ($\langle\eta\rangle$,
$\langle\overline{j}\rangle$) are obtained by averaging the microscopic
densities, this conclusion must extend to the macroscopic level. The
proper way of averaging is over an ensemble \cite{CHDEBJFG} with
a probability density $f({\overline{R}}_{kl},{\dot{\overline{R}}}_{kl};t)$
which depends on the positions of all charges and their velocities.
Alternatively the probability can be expressed as a function of the
atomic positions $\left({\overline{R}}_{k},{\dot{\overline{R}}}_{k}\right)$
and the internal coordinates $\left({\overline{r}}_{kl},{\dot{\overline{r}}}_{kl}\right)$.
Since the probability of finding the system in a particular configuration
is independent of the choice of the atomic reference point it follows
that: \begin{equation}
\begin{split}f({\overline{R}}_{k},{\dot{\overline{R}}}_{k},{\overline{r}}_{kl},{\dot{\overline{r}}}_{kl};t)={f}^{\prime}({{\overline{R}}^{\prime}}_{k},{{\dot{\overline{R}}}^{\prime}}_{k},{{\overline{r}}^{\prime}}_{kl},{{\dot{\overline{r}}}^{\prime}}_{kl};t)\end{split}
\label{eq:CHDDCIJF}\end{equation}

where we have already taken into account that the Jacobian of the
transformation (${\overline{R}}_{k}={{\overline{R}}^{\prime}}_{k}+\overline{d}$,
${\overline{r}}_{kl}={{\overline{r}}^{\prime}}_{kl}-\overline{d}$)
equals unity. If we take the sum over all building blocks of the first
contribution in (\ref{eq:CHDCICJA}) and take the average over this
distribution function, we find the unbound charge density $\rho$
in (\ref{eq:CHDGCJDC}): \begin{equation}
\begin{split}\rho=\langle\sum_{k}{q}_{k}\delta(\overline{r}-{\overline{R}}_{k})\rangle=\sum_{k}{q}_{k}\int\delta(\overline{r}-{\overline{R}}_{k}){f}_{k}({\overline{R}}_{k};t)d{\overline{R}}_{k}=\sum_{k}{q}_{k}{f}_{k}(\overline{r};t)\end{split}
\label{eq:CHDJFHFD}\end{equation}

where ${f}_{k}(\overline{r};t)$ is the probability density for finding
building block \emph{k} at $\overline{r}$. Again for simplicity we
consider now only the contribution of one particular kind of ions
and (\ref{eq:CHDJFHFD}) can then be written as: \begin{equation}
\begin{split}\rho=q{f}_{1}(\overline{r};t)\end{split}
\label{eq:}\end{equation}

where ${f}_{1}(\overline{r};t)$ is the probability density for finding
1 ion of this kind at $\overline{r}$. Note that (i) since we excluded
moving media $\frac{\partial\rho}{\partial t}=0$ and (ii) $\rho$
should be supplemented with a contribution from the free electrons;
but since these have no internal structure they are not relevant for
our subject.

If we choose another reference point (shift of the origin) then in
general another unbound charge density (due to the ions) is found:
\begin{equation}
\begin{split}{\rho}^{\prime}=q{{f}^{\prime}}_{1}(\overline{r};t)\end{split}
\label{eq:CHDIEHHI}\end{equation}

Due to (\ref{eq:CHDDCIJF}) the probability densities are related
by ${f}_{1}(\overline{r};t)={{f}^{\prime}}_{1}((\overline{r}-\overline{d});t)$
and therefore $\rho(\overline{r})={\rho}^{\prime}(\overline{r}-\overline{d})$.
After developing we find: \begin{equation}
\begin{split}\rho={\rho}^{\prime}-{d}_{i}{\partial}_{i}{\rho}^{\prime}+\frac{1}{2}{d}_{i}{d}_{j}{\partial}_{ij}{\rho}^{\prime}+\cdots\end{split}
\label{eq:CHDGBHGF}\end{equation}

This result is completely similar with that found for the corresponding
microscopic quantities where the $\delta$-functions in the latter
are here replaced by the probability density function. The same reasoning
can be followed for the higher order terms. For the polarization we
find: \begin{equation}
\begin{split}\overline{P}=\langle\sum_{k}{\overline{p}}_{k}\delta(\overline{r}-{\overline{R}}_{k})\rangle=\int\overline{p}({\overline{r}}_{l}){f}_{2}(\overline{r},{\overline{r}}_{l};t)d{\overline{r}}_{l}\end{split}
\label{eq:}\end{equation}

where the integration is over all internal coordinates and a similar
expression holds for the shifted origin where the distribution functions
are related by ${f}_{2}(\overline{r},{\overline{r}}_{l};t)={{f}^{\prime}}_{2}(\overline{r}-\overline{d},{{\overline{r}}^{\prime}}_{l};t)$.
Taking into account (\ref{eq:CHDDAFGG}) and (\ref{eq:CHDIEHHI})
we then find: \begin{equation}
\begin{split}{P}_{i}(\overline{r})={{P}^{\prime}}_{i}(\overline{r}-\overline{d})-{d}_{i}{\rho}^{\prime}(\overline{r}-\overline{d})\end{split}
\label{eq:CHDICDHE}\end{equation}

where we have also taken into account that the integration of ${{f}^{\prime}}_{2}$
over the internal coordinates yields ${{f}^{\prime}}_{1}$ and where
we have omitted the time dependence. After developing we find: \begin{equation}
\begin{split}{P}_{i}={{P}^{\prime}}_{i}-{d}_{i}{\rho}^{\prime}+{d}_{i}{d}_{j}{\partial}_{j}{\rho}^{\prime}-{d}_{j}{\partial}_{j}{{P}^{\prime}}_{i}+\cdots\end{split}
\label{eq:CHDCHHJH}\end{equation}

For the quadrupole density we find in the same way: \begin{equation}
\begin{split}{Q}_{ij}(\overline{r})={{Q}^{\prime}}_{ij}(\overline{r}-\overline{d})-\frac{1}{2}({d}_{i}{{P}^{\prime}}_{j}(\overline{r}-\overline{d})+{{P}^{\prime}}_{i}(\overline{r}-\overline{d}){d}_{j})+\frac{1}{2}{d}_{i}{d}_{j}{\rho}^{\prime}(\overline{r}-\overline{d})\end{split}
\label{eq:CHDIBAHA}\end{equation}

and then up to quadrupole order: \begin{equation}
\begin{split}{Q}_{ij}={{Q}^{\prime}}_{ij}-\frac{1}{2}({d}_{i}{{P}^{\prime}}_{j}+{{P}^{\prime}}_{i}{d}_{j})+\frac{1}{2}{d}_{i}{d}_{j}{\rho}^{\prime}+\cdots\end{split}
\label{eq:CHDFHGJB}\end{equation}

Although the individual multipole contributions do transform with
a shift of the origin, according to (\ref{eq:CHDGBHGF}) (\ref{eq:CHDCHHJH})
(\ref{eq:CHDFHGJB}) and so on, the total macroscopic charge density:
\begin{equation}
\begin{split}\langle\eta\rangle=\rho-{\partial}_{i}{P}_{i}+{\partial}_{ij}{Q}_{ij}+\cdots={\rho}^{\prime}-{\partial}_{i}{{P}^{\prime}}_{i}+{\partial}_{ij}{{Q}^{\prime}}_{ij}+\cdots\end{split}
\label{eq:CHDBHCIF}\end{equation}

does not, as can readily be verified. A similar conclusion can be
drawn with respect to the macroscopic current density $\langle\overline{j}\rangle$,
which is given by: \begin{equation}
\begin{split}\langle{j}_{i}\rangle={\partial}_{t}{P}_{i}-{\partial}_{t}{\partial}_{j}{Q}_{ji}+{e}_{ijk}{\partial}_{j}{M}_{k}={\partial}_{t}{{P}^{\prime}}_{i}-{\partial}_{t}{\partial}_{j}{{Q}^{\prime}}_{ji}+{e}_{ijk}{\partial}_{j}{{M}^{\prime}}_{k}\end{split}
\label{eq:CHDDEJDH}\end{equation}

where ${\partial}_{t}$ is the partial derivative with respect to
time and where the magnetization density transforms according to:
\begin{equation}
\begin{split}{M}_{i}(\overline{r})={{M}^{\prime}}_{i}(\overline{r}-\overline{d})-\frac{1}{2}{e}_{ijk}{d}_{j}{\partial}_{t}{{P}^{\prime}}_{k}(\overline{r}-\overline{d})\end{split}
\label{eq:CHDJFJBC}\end{equation}

and then up to quadrupole order: \begin{equation}
\begin{split}{M}_{i}={{M}^{\prime}}_{i}-\frac{1}{2}{e}_{ijk}{d}_{j}{\partial}_{t}{{P}^{\prime}}_{k}\end{split}
\label{eq:CHDDFBEJ}\end{equation}

Since at the macroscopic level only $\langle\eta\rangle,\langle\overline{j}\rangle$
have physical meaning, whereas in general $\rho,{P}_{i},{Q}_{ij},{M}_{i},\cdots$
do depend on the choice of reference point the latter must be considered
as equivalent representations of the same physical macroscopic charge
and current densities.

\section{Polarizabilities}

The constitutive equations of a medium express the response fields
${P}_{i},{Q}_{ij},{M}_{i},\cdots$ as a function of the fields ${E}_{i},{B}_{i}$.
This dependence can be local in space and time or more in general
the response fields can also depend on the values of the fields in
nearby points and/or in the past. We will use the same linear expressions
as in \cite{CHDFBJJA} (with slightly different notations) and we
will consider only a non magnetic medium and again limit ourselves
to terms of electric quadrupole/magnetic dipole order: \begin{equation}
\begin{split}{P}_{i}={P}_{i}^{\left(0\right)}+{\alpha}_{ij}{E}_{j}+{a}_{ijk}{\partial}_{j}{E}_{k}+{G}_{ij}{\partial}_{t}{B}_{j}\\
{Q}_{ij}={Q}_{ij}^{\left(0\right)}+{a}_{kij}{E}_{k}\\
{M}_{i}={M}_{i}^{\left(0\right)}-{G}_{ji}{\partial}_{t}{E}_{j}\end{split}
\label{eq:CHDFEFDG}\end{equation}

The values of the multipole densities in the absence of any fields
are indicated with a superscript $\left(0\right)$. In \cite{CHDFBJJA}
the origin dependence of the polarizabilities ${\alpha}_{ij},{a}_{ijk},{G}_{ij},\cdots$
have been found based on a microscopic (and quantum mechanical) theory.
Strictly speaking one should then consider first the local fields
in (\ref{eq:CHDFEFDG}) and then eliminate these so that only the
macroscopic fields remain. We avoid this complication by considering
(\ref{eq:CHDFEFDG}) as pure macroscopic equations where the polarizabilities
are phenomenological parameters. From the transformation properties
of the response fields which have been found in the previous section
we can deduce those of the polarizabilities.

Consider first the 2nd equation which we write in extenso as ${Q}_{ij}(\overline{r})={Q}_{ij}^{\left(0\right)}(\overline{r})+{a}_{kij}{E}_{k}(\overline{r})$.
A similar equation holds for the shifted reference point ${{Q}^{\prime}}_{ij}(\overline{r})={{Q}^{\prime}}_{ij}^{\left(0\right)}(\overline{r})+{{a}^{\prime}}_{kij}{E}_{k}(\overline{r})$.
In view of (\ref{eq:CHDIBAHA}) we combine these 2 into: \begin{equation}
\begin{split}{Q}_{ij}(\overline{r}+\overline{d})-{{Q}^{\prime}}_{ij}(\overline{r})={Q}_{ij}^{\left(0\right)}(\overline{r}+\overline{d})-{{Q}^{\prime}}_{ij}^{\left(0\right)}(\overline{r})+{a}_{kij}{E}_{k}(\overline{r}+\overline{d})-{{a}^{\prime}}_{kij}{E}_{k}(\overline{r})\end{split}
\label{eq:}\end{equation}

Using (\ref{eq:CHDIBAHA}) twice this becomes: \begin{equation}
\begin{split}-\frac{1}{2}{d}_{i}({{P}^{\prime}}_{j}(\overline{r})-{{P}^{\prime}}_{j}^{\left(0\right)}(\overline{r}))-\frac{1}{2}({{P}^{\prime}}_{i}(\overline{r})-{{P}^{\prime}}_{i}^{\left(0\right)}(\overline{r})){d}_{j}={a}_{kij}{E}_{k}(\overline{r}+\overline{d})-{{a}^{\prime}}_{kij}{E}_{k}(\overline{r})\end{split}
\label{eq:}\end{equation}

On the left side we use the first equation of (\ref{eq:CHDFEFDG})
only retaining terms up to quadrupole order (for the same reason the
$\overline{d}$ on the right side can be dropped) and we then find:
\begin{equation}
\begin{split}{a}_{kij}={{a}^{\prime}}_{kij}-\frac{1}{2}({d}_{i}{{\alpha}^{\prime}}_{jk}+{{\alpha}^{\prime}}_{ik}{d}_{j})\end{split}
\label{eq:CHDGAEAD}\end{equation}

In exactly the same way we find from the second equation in (\ref{eq:CHDFEFDG})
and using (\ref{eq:CHDJFJBC}): \begin{equation}
\begin{split}{G}_{ij}={{G}^{\prime}}_{ij}+\frac{1}{2}{e}_{jkl}{d}_{k}{{\alpha}^{\prime}}_{li}\end{split}
\label{eq:CHDHAGDA}\end{equation}

Lastly from the first equation in (\ref{eq:CHDFEFDG}) and using (\ref{eq:CHDICDHE})
we find at first: \begin{equation}
\begin{split}{\alpha}_{ij}{E}_{j}(\overline{r}+\overline{d})+{a}_{ijk}{\partial}_{j}{E}_{k}(\overline{r}+\overline{d})+{G}_{ij}{\partial}_{t}{B}_{j}(\overline{r}+\overline{d})={{\alpha}^{\prime}}_{ij}{E}_{j}(\overline{r})+{{a}^{\prime}}_{ijk}{\partial}_{j}{E}_{k}(\overline{r})+{{G}^{\prime}}_{ij}{\partial}_{t}{B}_{j}(\overline{r})\end{split}
\label{eq:CHDBIAHH}\end{equation}

Equating the terms of dipole-order on both sides one finds: \begin{equation}
\begin{split}{\alpha}_{ij}={{\alpha}^{\prime}}_{ij}\end{split}
\label{eq:CHDEGDDC}\end{equation}

All terms of quadrupole order cancel each other if one uses (\ref{eq:CHDGAEAD}),
(\ref{eq:CHDHAGDA}) and also takes into account Faraday’s law ${\partial}_{t}{B}_{j}(\overline{r})=-{e}_{jkl}{\partial}_{k}{E}_{l}(\overline{r})$
and the symmetry of ${\alpha}_{ij}$. The results (\ref{eq:CHDGAEAD})
(\ref{eq:CHDHAGDA}) and (\ref{eq:CHDEGDDC}) are exactly the same
as those found in \cite{CHDFBJJA}, although they have been derived
in a different way. Note that in transforming the polarizabilities
we have taken into account that with the shift of the origin (for
calculating the moments) one must also change the (position of the)
driving field (see e.g. (\ref{eq:CHDBIAHH}), where only the first
$\left(\overline{r}+\overline{d}\right)$ expression is relevant due
to our limitation to quadrupole order). It follows that if we use
the origin dependencies of the polarizabilities as used in \cite{CHDFBJJA}
and combine these with the proper shift of the field-point then the
resulting macroscopic charge and current densities do not depend on
the choice of origin as shown by (\ref{eq:CHDBHCIF}) and (\ref{eq:CHDDEJDH}).

\section{Reflection coefficients}

Turning now to the wave propagation in this medium it has been noted
\cite{CHDFBJJA} that the (plane wave) modes are origin-independent.
This was to be expected based on the analysis given above. In order
to determine the reflection coefficients for an air/medium interface
we need the boundary or jump conditions at this interface, taking
into account the quadrupole effects. Choosing the (\emph{x},\emph{y})-axes
into the plane of the interface and with the \emph{z}-axis pointing
into the medium these can be written as: \begin{equation}
\begin{split}{\epsilon}_{0}{\Delta}_{s}{E}_{x}={\partial}_{x}{\mathcal{P}}_{z}=-{\partial}_{x}{Q}_{zz}\\
{\epsilon}_{0}{\Delta}_{s}{E}_{y}={\partial}_{y}{\mathcal{P}}_{z}=-{\partial}_{y}{Q}_{zz}\end{split}
\label{eq:CHDFEGBG}\end{equation}
 \begin{equation}
\begin{split}{\epsilon}_{0}{\Delta}_{s}{E}_{z}={\pi}_{s}=-{P}_{z}+{\partial}_{x}{Q}_{zx}+{\partial}_{y}{Q}_{zy}+{\partial}_{j}{Q}_{jz}\end{split}
\label{eq:CHDIBIHA}\end{equation}
 \begin{equation}
\begin{split}\frac{{\Delta}_{s}{B}_{x}}{{\mu}_{0}}={-K}_{y}=-{M}_{x}+{\partial}_{t}{Q}_{zy}\\
\frac{{\Delta}_{s}{B}_{y}}{{\mu}_{0}}={K}_{x}=-{M}_{y}-{\partial}_{t}{Q}_{zx}\end{split}
\label{eq:CHDDFIFA}\end{equation}
 \begin{equation}
\begin{split}{\Delta}_{s}{B}_{z}=0\end{split}
\label{eq:CHDCEFAA}\end{equation}

where ${\Delta}_{s}$ stands for the field on the air side minus the
field in the medium. ${\pi}_{s}$ is the (bound) surface charge density,
${K}_{x},{K}_{y}$ are the components of the (bound) surface current
density and ${\mathcal{P}}_{x},{\mathcal{P}}_{y},{\mathcal{P}}_{z}$
those of the surface polarization density. These equations are the
same as those used in \cite{CHDFBJJA} except for (\ref{eq:CHDIBIHA})
where the contribution ${\partial}_{z}{Q}_{zz}$ in the last term
on the RHS is missing in \cite{CHDFBJJA} (and the conditions differ
from those published previously \cite{CHDJEGJD}). The quadrupole
contributions in (\ref{eq:CHDFEGBG})-(\ref{eq:CHDDFIFA}) can be
understood as follows: in the bulk of the medium the quadrupole density
is equivalent with a polarization density ${P}_{i}^{Q}=-{\partial}_{j}{Q}_{ji}$.
At the interface this density becomes singular and gives rise to a
surface polarization density ${\mathcal{P}}_{i}=-{Q}_{zi}$. The normal
component ${\mathcal{P}}_{z}=-{Q}_{zz}$ gives rise to a local voltage
difference over the boundary and its variation along the boundary
enters Faraday’s law and explains (\ref{eq:CHDFEGBG}). The in-plane
components on the other hand give rise to a surface charge density
$-{\partial}_{x}{\mathcal{P}}_{x}-{\partial}_{y}{\mathcal{P}}_{y}$,
which explains the first 2 quadrupole contributions in (\ref{eq:CHDIBIHA}),
and a surface current density with components ${\partial}_{t}{\mathcal{P}}_{x}$
and ${\partial}_{t}{\mathcal{P}}_{y}$, which explain the contributions
in (\ref{eq:CHDDFIFA}). In the bulk the polarization ${P}_{i}^{Q}=-{\partial}_{j}{Q}_{ji}$
is equivalent with a charge density $-{\partial}_{i}{P}_{i}^{Q}={\partial}_{i}{\partial}_{j}{Q}_{ji}$
and a current density ${\partial}_{t}{P}_{i}^{Q}=-{\partial}_{t}{\partial}_{j}{Q}_{ji}$.
The former has a singularity at the surface leading to a surface charge
density ${\partial}_{j}{Q}_{jz}$, which is the last contribution
in (\ref{eq:CHDIBIHA}). Note that without this term the normal component
of the accompanying current density is not balanced at the surface.

Using Faraday’s law on both sides of the interface it is easily
seen that (\ref{eq:CHDFEGBG}) implies (\ref{eq:CHDCEFAA}). Similarly
using instead Ampère’s law one can also verify that (\ref{eq:CHDDFIFA})
implies (\ref{eq:CHDIBIHA}). Therefore we only need the equations
(\ref{eq:CHDFEGBG}) and (\ref{eq:CHDDFIFA}) to find the reflection
coefficients. In general the quadrupole contributions in these equations
will change if another reference point is chosen, as in the bulk of
the material. However just as in the bulk this change will be balanced
exactly by the displacement of the (lower order) dipole moment and
therefore if calculated correctly no change in reflection coefficients
should be found. However whereas in the bulk the latter effect is
accounted for automatically (and yields an origin independent wave
equation and origin independent modes) at the free surface it must
be taken into account explicitly. It seems indeed logical that when
moving the (atomic) reference point, and therefore all multipoles,
over a distance $\overline{d}$, then one should at the macroscopic
level, move the boundary surface over the same distance with the immediate
conclusion that only ${d}_{z}$ will be relevant. The effect of such
a shift (alone) on the reflection coefficients is easily found to
first order as: \begin{equation}
\begin{split}{\left(dR\right)}_{\text{ss}}=-2j{k}_{z}{d}_{z}R\end{split}
\label{eq:CHDIADAA}\end{equation}

where “ss” stands for “surface shift” and with ${k}_{z}$
the normal wave vector component in free space (we use the time dependence
$e^{j\omega t}$). If the reflection coefficients are origin independent
then the change in the reflection coefficients due to the change of
the quadrupole moments alone must exactly be opposite to the one in
(\ref{eq:CHDIADAA}): \begin{equation}
\begin{split}{\left(dR\right)}_{\text{qc}}=2j{k}_{z}{d}_{z}R\end{split}
\label{eq:CHDIAFJE}\end{equation}

where “qc” stands for “quadrupole change” and this should
be true irrespective of any symmetry properties of the medium. Proving
this relation by direct analytical calculation turns out too complex
so far in the most general case, this means for oblique incidence
on an arbitrary medium%
\footnote{“arbitrary” means in this context “without any special symmetry
properties”.%
}. However we could prove relation (\ref{eq:CHDIAFJE}) if we relax
one of these conditions: (i) perpendicular incidence on an arbitrary
medium or (ii) oblique incidence on a medium with a 4-fold symmetry
axis perpendicular to the surface. Further evidence for the general
validity of (\ref{eq:CHDIAFJE}) can be gathered directly from the
relevant boundary conditions (\ref{eq:CHDFEGBG}) and (\ref{eq:CHDDFIFA}).
Using (\ref{eq:CHDFHGJB}) and (\ref{eq:CHDDFBEJ}) the changes in
${\mathcal{P}}_{z},{K}_{x},{K}_{y}$ due to the change of the quadrupole
moment are given by%
\footnote{Note that due to our definition of $\overline{d}$ ($\overline{R}={\overline{R}}^{\prime}+\overline{d}$)
the difference equals unaccented quantities minus accented quantities.%
}: \begin{equation}
\begin{split}{\left(d{\mathcal{P}}_{z}\right)}_{\text{qc}}={d}_{z}{P}_{z}\\
{\left(d{K}_{x}\right)}_{\text{qc}}={d}_{z}{\partial}_{t}{P}_{x}\\
{\left(d{K}_{y}\right)}_{\text{qc}}={d}_{z}{\partial}_{t}{P}_{y}\end{split}
\label{eq:}\end{equation}

Since the bulk current density has components ${\partial}_{t}{P}_{x},{\partial}_{t}{P}_{y}$
along the free surface these variations are indeed as if the surface
moved outwards over ${d}_{z}$ whereas ${d}_{z}>0$ actually represents
an inward movement of the surface, since the \emph{z}-axis points
into the medium. Finally we note that the lowest order term shown
in (\ref{eq:CHDIAFJE}) is imaginary since it corresponds with a pure
phase shift. In \cite{CHDFBJJA} the variation of the modulus of \emph{R}
only was considered and therefore the relevant lowest order contribution
to $dR$ was missed. The variation of the modulus is indeed of a higher
than quadrupole order and since calculations have been done up to
quadrupole order only this is a meaningless result.

\section{Transformations}

In the frequency-domain the constitutive equations can generally be
expressed as: \begin{equation}
\begin{split}{D}_{i}={\epsilon}_{ij}{E}_{j}+{\gamma}_{ij}{B}_{j}\\
{H}_{i}={\beta}_{ij}{E}_{j}+{\nu}_{ij}{B}_{j}\end{split}
\label{eq:CHDCAEHJ}\end{equation}

where the material constants can be calculated from (\ref{eq:CHDFEFDG}):
\begin{equation}
\begin{split}{\epsilon}_{ij}={\epsilon}_{0}{\delta}_{ij}+{\alpha}_{ij}+j{k}_{m}({a}_{jmi}-{a}_{imj})\\
{\gamma}_{ij}=j\omega{G}_{ij}\phantom{\rule{10mm}{0ex}}{\beta}_{ij}=j\omega{G}_{ji}\phantom{\rule{10mm}{0ex}}{\nu}_{ij}={\mu}_{0}^{-1}{\delta}_{ij}\end{split}
\label{eq:CHDJHBFB}\end{equation}

Due to (\ref{eq:CHDGAEAD}), (\ref{eq:CHDHAGDA}) these material constants
will usually be “origin dependent” and therefore at first sight
also material dependent properties like e.g. reflection coefficients.
As explained in the previous section the change of reference point
should be accompanied by a shift of the free surface boundary and
when both effects are taking into account the reflection coefficients
are invariant. Raab and De Lange\cite{CHDFBJJA} \cite{CHDBJBCB}
\cite{CHDCFCBH} did not take the latter effect into account but instead,
in order to get rid of the “origin dependence” of the material
constants in (\ref{eq:CHDJHBFB}), they developed a “transformation
theory”, which essentially applies changes $\Delta{G}_{ij}$ and
$\Delta{a}_{ijk}$ to the polarizabilities in such a way that Maxwell’s
equations remain invariant. The new formulation is supposed to be
equivalent with the original one, but since both formulations lead
to different physical results (in casu different reflection coefficients)
this cannot be true. This is due to the fact that no attention was
given to the boundary conditions. In fact it is almost inescapable
that different reflection coefficients (in casu origin independent
ones) could only have been obtained using this procedure by changing
the boundary conditions.

In order for the boundary conditions (\ref{eq:CHDFEGBG}) and (\ref{eq:CHDDFIFA})
to be invariant the RHS terms should not change by the transformation,
since the fields $\overline{E}$ and $\overline{B}$ are unique, thus:
\begin{equation}
\begin{split}\Delta{Q}_{zz}=0\phantom{\rule{10mm}{0ex}}\Delta{K}_{x}=0\phantom{\rule{10mm}{0ex}}\Delta{K}_{y}=0\end{split}
\label{eq:}\end{equation}

With (\ref{eq:CHDFEFDG}) and since these conditions should hold for
an arbitrary electric field it follows that: \begin{equation}
\begin{split}\Delta{a}_{jzz}=0\\
-\Delta{G}_{jy}+\Delta{a}_{jzx}=0\\
\Delta{G}_{jx}+\Delta{a}_{jzy}=0\end{split}
\label{eq:}\end{equation}

for any \emph{j}. Since in addition boundary conditions should remain
invariant along any possible boundary (not only the $\left(x,y\right)$-plane
considered in this example) these conditions should also hold after
cyclic permutation $x\rightarrow y\rightarrow z$. It then readily
follows that $\Delta{G}_{ij}=0$ and $\Delta{a}_{ijk}=0$ meaning
that no transformation leaves the boundary conditions invariant.

\section{Conclusions}

Changing the point of reference for calculating the multipole moments
for the building blocks of a medium has two consequences. The moments
themselves usually change with the reference point and the position
in space of these elementary multipoles changes. If both effects are
taken into account then the resulting charge and current densities
are independent of the reference point. This has been shown at the
microscopic and at the macroscopic level. The same interpretation
can be applied to e.g. the reflection coefficients for a semi infinite
medium: if with a change of origin the boundary of the medium is shifted
accordingly, then the reflection coefficients are invariant. If the
shift of the boundary is not taken into account then the reflection
coefficient will show a variation corresponding with a pure phase
shift. There is no need to remedy this “origin dependence” effect
with a “transformation theory”. In fact there is no transformation
which leaves all possible boundary conditions invariant. We have performed
all calculations including the multipoles of electric quadrupole/magnetic
dipole order and for a non-magnetic and non-absorbing medium but we
have no doubt that these results are still valid for more general
materials.

\end{document}